# Spin-Orbital Entangled Liquid State in the Copper Oxide $Ba_3CuSb_2O_9$

Huiyuan Man[1], Mario Halim[1], Hiroshi Sawa[2], Masayuki Hagiwara[3], Yusuke Wakabayashi[4], Satoru Nakatsuji[1,5,*]

[1] *Institute for Solid State Physics, The University of Tokyo, Kashiwa, Chiba 277-8581, Japan*

[2] *Department of Applied Physics, Graduate School of Engineering, Nagoya University, Nagoya 464-8603, Japan*

[3] *Center for Advanced High Magnetic Field Science, Graduate School of Science, Osaka University, Toyonaka, Osaka 560-0043, Japan*

[4] *Division of Materials Physics, Graduate School of Engineering Science, Osaka University, Toyonaka, Osaka 560-8531, Japan*

[5] *CREST, Japan Science and Technology Agency (JST), 4-1-8 Honcho Kawaguchi, Saitama 332-0012, Japan*

Corresponding author: Satoru Nakatsuji (satoru@issp.u-tokyo.ac.jp)

**Abstract**

Structure with orbital degeneracy is unstable toward spontaneous distortion. Such orbital correlation usually has a much higher energy scale than spins, and therefore, magnetic transition takes place at a much lower temperature, almost independently from orbital ordering. However, when the energy scales of orbitals and spins meet, there is a possibility of spin-orbital entanglement that would stabilize novel ground state such as spin-orbital liquid and random singlet state. Here we review on such a novel spin-orbital magnetism found in the hexagonal perovskite oxide $Ba_3CuSb_2O_9$, which hosts a self-organized honeycomblike short-range order of a strong Jahn-Teller ion $Cu^{2+}$. Comprehensive structural and magnetic measurements have revealed that the system has neither magnetic nor Jahn-Teller transition down to the lowest temperatures, and Cu spins and orbitals retain the hexagonal symmetry and paramagnetic state. Various macroscopic and microscopic measurements all indicate that spins and orbitals remain fluctuating down to low temperatures without freezing, forming a spin-orbital entangled liquid state.

**Introduction**

At high temperatures, electronic orbitals are degenerate in energy and maintain its spherical shape as thermal fluctuations dominate anisotropic interactions. In a lattice, however, this degeneracy is usually lifted by spontaneous symmetry breaking due to anisotropic interaction inherent to the strong Coulomb energy [1]. In spin systems, the possibility of a spin liquid state without symmetry breaking has been proposed for geometrically frustrated magnets, which has now become one of major research topics in condensed matter physics [2, 3]. On the other hand, electronic orbital degree of freedom, which responds to "the crystal electric field", is the minimal unit of degrees of freedom related to "the shape" of localized valence electron inside a material, and is an important element that determines, based on the Neumann principle [4], the relation between anisotropic physical properties and crystallographic point group, and the dimensionality of the system responding to the external field. "Orbital liquid" material, which does not show any orbital freezing at low temperatures, has been predicted theoretically since 1990s and its experimental verification in some transition metal oxides have been carried out [5-12]. However, their realization has remained hypothetical to date.

**Various Attempts to Realize an Orbital Liquid State**

It should be noted that it is far more difficult to realize an orbital liquid state than a spin liquid as orbitals have stronger anisotropic characters, higher energy scales and thus, weaker quantum effects than spin systems in general [1]. Hence, relevant choice of system, precise physical properties measurements and in-depth theoretical understanding are necessary for the search of the candidate materials for orbital liquids. Among such attempts to realize the novel state, the quasi two dimensional (2D) triangular lattice magnet $Li_{1-x}Ni_{1+x}O_2$ with active $e_g$ orbitals was once thought to be a promising candidate [9, 11, 13, 14]. Extensive research has been made to reveal the origin of the observed behaviors hinted a possible orbital liquid state. In contrast with the general wisdom that the intersite orbital exchange coupling should be highly anisotropic due to the Hund's coupling as well as anisotropic character of the wave function, an interesting limit of the spin-orbital lattice (Kugel-Khomskii) model with highly isotropic SU(4) symmetry was discovered and found to stabilize a spin-orbital liquid state [7]. More realistic models have been proposed to understand experimental observations revealing intrinsic and extrinsic ferromagnetic interactions supported by orbital ordered and/or disordered states [9, 11, 13, 14]. Experimentally, however, it was later found that $Li_{1-x}Ni_{1+x}O_2$ exhibits orbital freezing [10].

Theoretical studies have been made for the transition metal oxides with the $t_{2g}$ orbital degeneracy as well and found rich quantum states including various spin-orbital liquid states stabilized by the enhanced quantum fluctuations due to the frustrated spin and orbital interactions and spin-orbit coupling [8, 15-17]. Among them, significant attention has been paid to a theory on spin and orbital states in the Mott insulator $LaTiO_3$ that argued that orbital liquid could be realized as a quantum coherent state lifting orbital degeneracy due to geometrical frustration [8]. However, the later X-ray and neutron-diffraction as well as thermal transport studies indicate that $t_{2g}$ orbitals becomes non-degenerate due to the Jahn-Teller effect at low temperatures that competes against orbital liquid state at high temperatures [18, 19]. In addition, possibility of the spin orbital entanglement due to the combination of electron correlation and spin-orbit coupling was proposed to explain the observed signature of the orbital liquid state in the spinel lattice compound $FeSc_2S_4$ [20-23]. However, later it was showed that this material also exhibits orbital freezing at low temperatures [12]. Thus, up to now there has not been any established example of a compound realizing an orbital liquid state.

Recently, we found that the copper oxide material, 6H-$Ba_3CuSb_2O_9$ (from hereafter BCSO) is the first dense copper oxide system that is free from the Jahn-Teller (JT) transition down to low temperatures [24]. There are two kinds of samples in this system: one is the hexagonal sample that maintains the hexagonal structure down to low temperatures, and the other is the orthorhombic sample that undergoes a phase transition to orthorhombic structure with an orbital order. They have very similar structure at room temperature; detail will be discussed later. For the single crystal of the hexagonal sample, there are three-fold axes on Cu-site and no static JT distortion due to oxygen displacement is observed down to low temperatures. Because the energy gained from JT distortion is normally of the order of several thousand Kelvin, it is hard to imagine that the orbital retains its spherical shape below room temperature. Our recent experimental studies indicate that this system has an orbital liquid state, where the orbital states with three different quantized axes are quantum mechanically entangled. Considering the dynamical Jahn-Teller effect, the competition between intersite "spin-orbital singlet dimer" and intrasite "local JT singlet" has been discussed using a quantum-dimer model derived from the Hamiltonian hosting both the superexchange interaction and the dynamical JT effect. "spin-orbital singlet dimer" means two spins in a nearest-neighbor bond form a singlet state associated with an orbital polarization along the bond, and "local JT singlet" indicates that an orbital polarization is quenched due to the dynamical JT effect, as shown in Figure 1 (a) and (b) [25, 26]. The theoretical work has proposed that the above competition leads to a spin-orbital resonant state where dynamical resonant valence bonds are realized with entangled local spin-singlet states and parallel orbital configurations [25], as schematically shown in Figure 1 (c). Another theory by P. Corboz *et al*. raises the possibility of a quantum spin-orbital liquid on the honeycomb lattice in BCSO using SU(4) symmetric Kugel-Khomskii model [27]. Decoupled spin chains were theoretically proposed to understand the measured magnetic susceptibility of BCSO [28]. In this paper, we will review the current research status on the spin-orbital liquid state in BCSO.

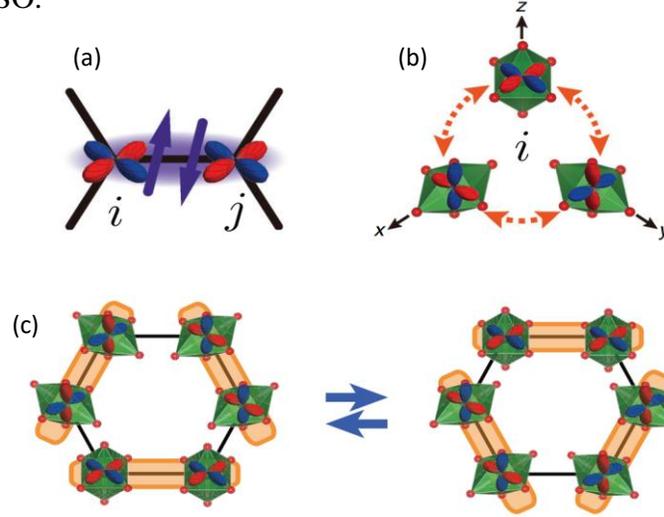

Figure 1. Schematic views of (a) a spin-orbital singlet dimer, (b) a JT singlet, and (c) a spin-orbital resonant state. The spin-singlet and parallel-orbital bonds are denoted by the shaded bonds. Figures are taken from ref [25] and ref [26].

**Geometrical Frustration in Crystal Structure**

- **6H-Perovskite $Ba_3CuSb_2O_9$**

The structure of 6H-$Ba_3CuSb_2O_9$ was first reported in 1978 as a hexagonal perovskite compound [29]. Generally, a perovskite-type structure is represented by the formula $ABX_3$. By rewriting the formula as $Ba(Cu_{1/3}Sb_{2/3})O_3$, one notices that A site is occupied by Ba, while B site is shared by 1/3 of Cu and 2/3 of Sb, forming a type of the perovskite structure. The valence of Cu and Sb can be estimated to be 2+ and 5+, respectively, because Barium and Oxygen valences should be 2+ and 2-, and thus the remaining $(Cu_{1/3}Sb_{2/3})$ valence must be 4+. Then, electronic configuration of $Sb^{5+}$ should be $4d^{10}$ (non-magnetic) while $Cu^{2+}$ has the $3d^9$ ($S = 1/2$) state. Generally, in the perovskite-type compounds, $BX_6$ octahedra may be connected by sharing corners ($SrTiO_3$, *etc*.), or sharing faces ($BaNiO_3$, *etc*.), depending on the size of A-site ions. In the case of 6H-$Ba_3CuSb_2O_9$, $BX_6$ octahedra are connected by sharing both corners and faces. According to the literature [29], the corner sharing octahedra $BX_6$ is occupied only by Sb (Figure 2). On the other hand, on the B-site of the face sharing octahedra, Cu and Sb form a pair and create $CuSbO_9$ "dumbbell structure". This structure carries an electric dipole moment due to the charge difference between $Cu^{2+}$ and $Sb^{5+}$. If the non-centrosymmetric space group P6$_3$mc is chosen as reported in 1978 [29], there should be a three-fold rotation axis on this dumbbell. By symmetry, this three-fold rotation axis guarantees the degeneracy of the $e_g$ orbitals of 3d electron, and thus the orbital degree of freedom of Cu$^{2+}$ ion.

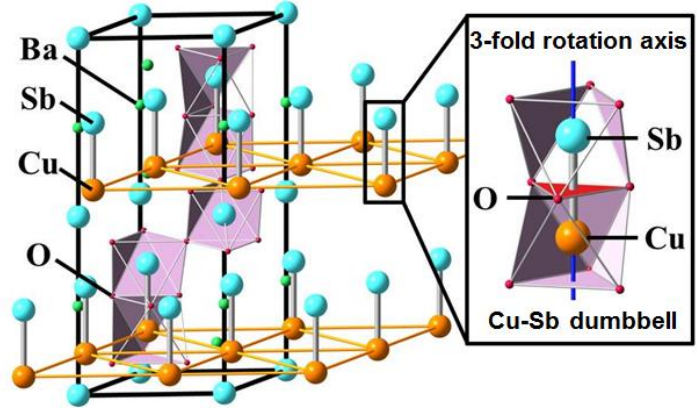

Figure 2. Crystal structure of Ba$_3$CuSb$_2$O$_9$ reported in 1978 [29], which is later found inconsistent with the X-ray diffraction results [24]. Cu-Sb dumbbells align themselves in the same direction, which forms a two-dimensional triangular lattice of Cu ions.

Interestingly, the Cu-sites form two-dimensional triangular lattice in this non-centrosymmetric structure. On the other hand, the Heisenberg quantum spins on a triangular lattice antiferromagnet is known to have strong magnetic frustration and thus the exploration of the low temperature magnetism defines an interesting subject. Therefore, we have performed the muon spin resonance and magnetothermal measurements, and found the absence of a long range magnetic order down to the lowest temperatures of 20 mK [24]. Similar measurements had been performed first by H. D. Zhou *et al.* [30], which raised interesting possibility that this system should have a quantum spin liquid state on the triangular lattice [30, 31]. Strikingly, however, our further structural analyses [24] have revealed a significantly different space group from the previous reports [29, 30], in which Cu does not realize triangular lattice. First, let us explain the crystal structure of this system at room temperature.

- **Determination of structural space group**

Generally, the reflection condition derived from the X-ray diffraction provides the information on centering, glide, and screw symmetry operations as well as the Bravais lattice. However, the information on centrosymmetry, mirror planes, rotation and inversion axes cannot be determined based on the reflection conditions alone. Therefore, we need to select the best space group by comparing with the measured intensity and the calculated intensity using various structure models.

| h k l | $|F_{obs}|$ | $|F_{cal\_P63mc}|$ | $|F_{cal\_P63/mmc}|$ |
|---|---|---|---|
| 0 0 2 | 6.24 | 29.55 | 5.36 |
| 0 0 4 | 53.00 | 56.77 | 56.71 |
| 0 0 6 | 110.54 | 107.84 | 108.70 |

Table 1. Comparison between the calculated crystal structure factor of 00*l* reflection, using *P*6$_3$*mc* and *P*6$_3$/*mmc* space group, and the observed value.

First, we performed the single crystal X-ray diffraction on the beam line (BL02B2) at the synchrotron radiation facility SPring-8, and confirmed that crystal structure is hexagonal with Laue group of 6/*mmm*. We also found that the reflection condition is $l = 2n$ (*n*: integer) for *hhl*. This condition allows us to limit the space group to either *P*6$_3$*mc*, $P\bar{6}2c$, or *P*6$_3$/*mmc*. Among them, the space group $P\bar{6}2c$, which has no screw symmetry, is incompatible with 6H-perovskite structure, and therefore is excluded. Von. P. Köhl actually has chosen the non-centrosymmetric space group of *P*6$_3$*mc* among the two remaining candidates *P*6$_3$*mc* and *P*6$_3$/*mmc* [29, 30, 32]. Next, from the obtained diffraction intensities, we estimated the structure factor $|F_{obs}|$ from experimental observations, and compared it with the structure factor calculated for space group *P*6$_3$*mc* ($|F_{cal\_P63mc}|$). The calculation results for several *00l* peaks are shown in Table 1. Clear discrepancy is obtained in 002 peak reflection intensity. Namely, the *P*6$_3$*mc* space group failed to reproduce the observed intensity. Then we performed similar procedure for space group *P*6$_3$/*mmc*. The space group *P*6$_3$/*mmc* requires a mirror plane on the oxygen plane between two CuSbO$_9$ octahedra (Cu-Sb dumbbell), and thus indicates that both Cu and Sb sites are crystallographically equivalent. In other words, the occupancy of Cu and Sb for the metal sites in CuSbO$_9$ bioctahedra is both 50%.

As shown in Figure 3, *P*6$_3$/*mmc* much better reproduces peak intensity than *P*6$_3$*mc*. Through examining local structure around Cu and Sb based on the analysis of EXAFS results, we found that more than 95% of the dumbbells are Cu-Sb pairs, and there is almost no Cu-Cu or Sb-Sb dumbbell [24]. Having noticed a new space group candidate, we examined it further by measuring the dielectric properties. As mentioned above, the Cu-Sb dumbbell has a large electric dipole moment due to the valence difference between $Cu^{2+}$ and $Sb^{5+}$. Therefore, macroscopic pyroelectricity is expected for the structure with *P*6$_3$*mc* symmetry where the Cu-Sb dumbbells form a ferroic arrangement. No pyroelectricity was observed for the volume of 1μm × 1μm × 5μm [24]. This result strongly supports the mirror symmetry parallel to the *c*-plane. Figure 4 shows the results of the dielectric constant measurement performed below room temperature. Neither large dielectric coefficient nor anomalies related to a phase transition is observed, consistent with the *P*6$_3$/*mmc* symmetry without pyroelectricity.

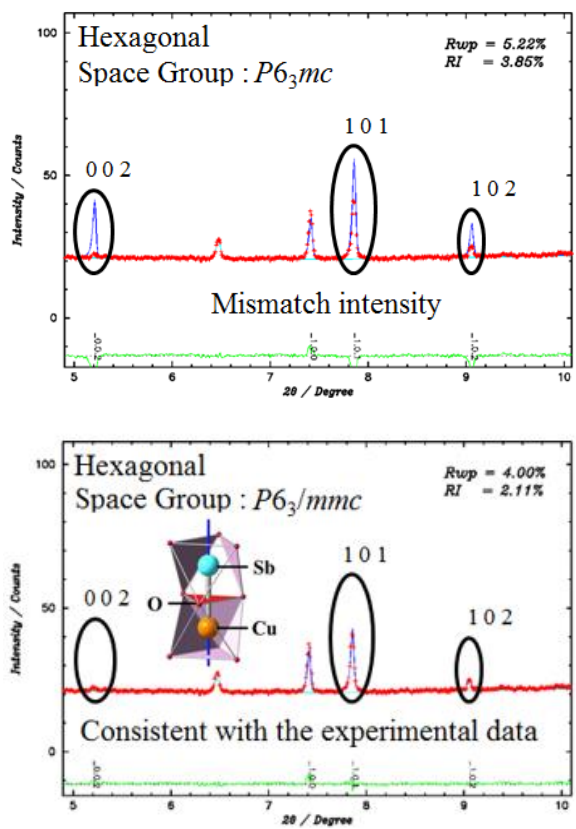

Figure 3. Analysis results for the two space groups and the powder XRD pattern measured at SPring-8 BL02B2. Red line is the measured data. The blue line is calculated intensity. Green line represents the residuals. *Rwp* and *RI* are the refinement parameters. The inset represents the Cu-Sb dumbbell. Same results are obtained for both hexagonal and orthorhombic samples at room temperature.

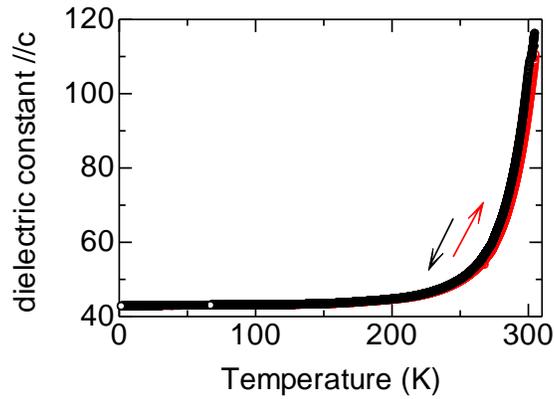

Figure 4. Dielectric constant along the c-axis as a function of temperature for a single crystal of the orthorhombic phase of $Ba_3CuSb_2O_9$. The measurement frequency is 1 kHz.

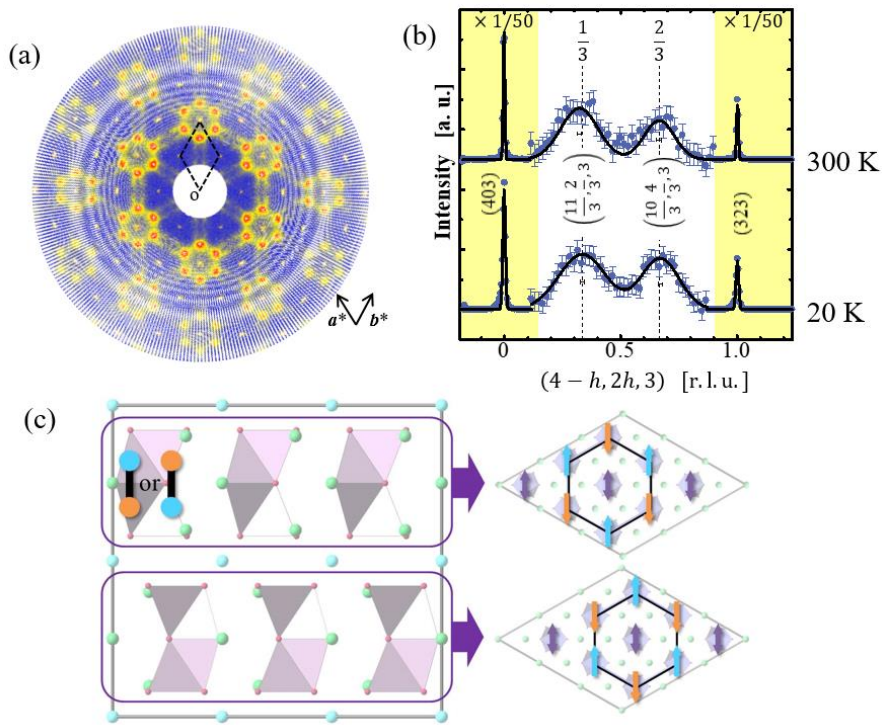

Figure 5. X-ray diffuse scattering and 3×3 times expanded unit cell for an orthorhombic single-crystalline sample. (a) Diffuse scattering appears at reciprocal lattice plane of $hk10$. $a^*$ and $b^*$ are the directions in reciprocal space. (b) Diffraction profile of diffuse scattering along (4-$h$, 2$h$, 3) at 300 K and 20 K. The intensity of (403) and (323) is divided by 50 in order to show the whole peaks. The dashed lines show $h=1/3$ and $2/3$. The diffusion scattering patterns are almost the same regardless of whether or not it shows phase transition at low temperatures and does not change in temperature. (c) Schematic of each dumbbell orientation on two Cu-Sb dumbbell layers from the included unit cell. Arrows indicate dumbbell orientation.

- **Geometrical frustration and short-range order of Cu-Sb dumbbell**

The valence of $CuSbO_9$ can be obtained as $[(Cu^{2+})(Sb^{5+})(O^{2-})_9]^{11-}$ and electronic state for $Cu^{2+}$ and $Sb^{5+}$ should be regarded as $[Ar]3d^9$ and $[Kr]4d^{10}$, respectively. In the structure with $P6_3/mmc$ symmetry, both Cu and Sb sites are equivalent. While the Cu-Sb dumbbell forms a triangular lattice, Cu and Sb ions occupy randomly the dumbbell sites and do not support the formation of a two-dimensional triangular lattice of the $Cu^{2+}$ $S = 1/2$ spin. We observed strong diffuse scattering from single crystal X-ray diffraction experiments around the reciprocal lattice points of (⅓ ⅓ $l$) ($l$ is integer) (Figures 5a and 5b). Since there is no temperature dependence in both diffuse scattering intensity and full width half maximum (Figure 5b), the structural short-range order should be stabilized in the crystal growth process. We confirmed that this diffuse scattering originates from the short-range order of the Cu-Sb dumbbell orientation. In other words, randomness is determined solely by the orientation of the dumbbell. This indicates that each Cu-Sb dumbbell orientation affects its neighbor's dumbbell direction through antiferroic coupling, leading to a macroscopic antiferroelectric order.

Figure 5c illustrates the short-range order of the Cu-Sb dumbbell. The unit cell is now enlarged to include two stacked dumbbell layers along the c-axis. On each dumbbell layer there are nine dumbbells, forming the two-dimensional triangular lattice. If we assume that 6 dumbbells per layer form an antiferroelectric arrangement, there remain 3 dumbbells per layer with undetermined orientation. These three dumbbells are free to pick any orientation, because all possible arrangements are energetically equivalent when we take into account solely the nearest neighbor interaction. Significantly, this situation, where dumbbell orientation is geometrically frustrated, is equivalent to the geometrical frustration of Ising spins on the triangular lattice [33]. There are $2^6$=64 combinations for upper and lower layer (total 6) dumbbell orientation. However, only one solution where all six dumbbells have the same orientation can reproduce the reflection condition of the diffuse intensity corresponding to the 3×3×1 unit cell. Namely, when connecting neighboring Cu sites, a short-range order honeycomb structure should be formed on the 2D triangular lattice with Ising AF interaction. However, this structure cannot eliminate the frustration from the electric dipoles. The Cu-Sb dumbbells are arranged in a ferrielectric form within the domain of this short-range order. Thus, the ferroelectricity derived from the electric dipole moment of the Cu-Sb dumbbell should appear in the microscopic region. However, since macroscopic pyroelectricity was not detected in the dielectric constant measurements, we expect that the ferrimagnetic domains of the honeycomb structure cancel the polarization with each other. The Cu sites in this system locally form both the honeycomb lattice and the triangular lattice.

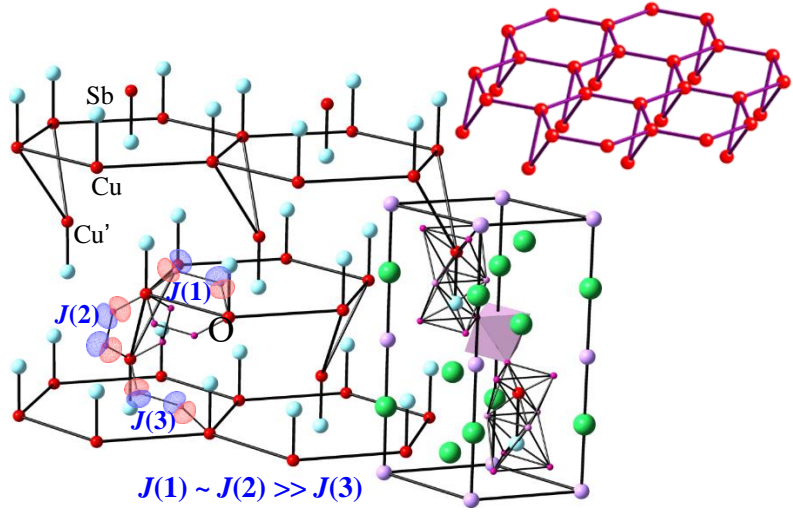

Figure 6. $Ba_3CuSb_2O_9$ crystal structure confirmed by synchrotron X-ray diffraction. Cu ion (red) forms short-range order which resembles to honeycomb lattice. The interaction of Cu-Cu sites, Cu-Cu' sites, and nearest neighbor $Cu^{2+}$ are indicated by J(1), J(2) and J(3). Upper right figure shows the structure composed only of Cu ions.

Figure 6 shows the idealized crystal structure based on the above discussion. This structure is not simply the honeycomb lattice structure of the Cu sites, but also contains the Cu' sites whose magnetic interaction cannot be omitted. Namely, the quantum chemistry type analyses reveal that both the magnetic interaction $J_1$ connecting the neighboring Cu-Cu sites within the honeycomb lattice and the coupling $J_2$ between Cu-Cu' have the super-exchange interaction pathways using the Cu-O-O-Cu route, indicating that the interaction strength is comparable for both $J_1$ and $J_2$, and the Cu' sites cannot be ignored. On the other hand, the coherence length calculated from the diffuse scattering peak is only about 10 Å (Figure 5b), roughly corresponding to the diagonal length of one honeycomb lattice unit. Thus, the actual structure may well have a depleted honeycomb lattice of the Cu site. In fact, the later experiment and theory indeed confirm that the Cu sites form rather local stripe order with a part of honeycomb lattice locally decorated by the Cu' sites, as we discuss in the following.

Short-range order of Cu-Sb dumbbell orientation is further examined based on the fluorescence X-ray holography and X-ray diffuse scattering characterized by the modulation wave vector close to (1/3 1/3 0) [34]. The diffuse intensity was confirmed to reflect the short-range order of Cu-Sb dumbbell orientation, while the majority of the intensity is originated from the JT distortion of the $CuO_6$ octahedra. In accordance with the EXAFS result [24], the amount of JT distortion is independent of temperature as the diffuse intensity shows no temperature variation. The diffuse intensity is insensitive to the orientation of the JT distortion, very different from the diffuse scattering around the Bragg reflections discussed in the later section. Pair distribution function (PDF) derived from the measured diffuse intensity around (1/3 1/3 0) away from reciprocal lattice points are presented in Figure 7 (a). The PDF shows strong negative correlation at *a*, the same amount of positive correlation at *a*+2*b*, and some weak other correlations. Panel (b) shows the PDF for the cluster presented in the inset, which reproduces the experimental observations well. Note that the cluster is very similar to the structure presented in Figure 5 (c) and Figure 6. Keep in mind that it is a cluster, not the unit cell. No clear periodicity is found in Cu-Sb orientation. A real space Cu-Sb arrangement providing such a PDF is presented in the panel (c). It has no clear periodicity, while it has some degree of tendency to have different kinds of element to come to the first neighbor, and locally resembles but not exactly the structure shown in Figure 6.

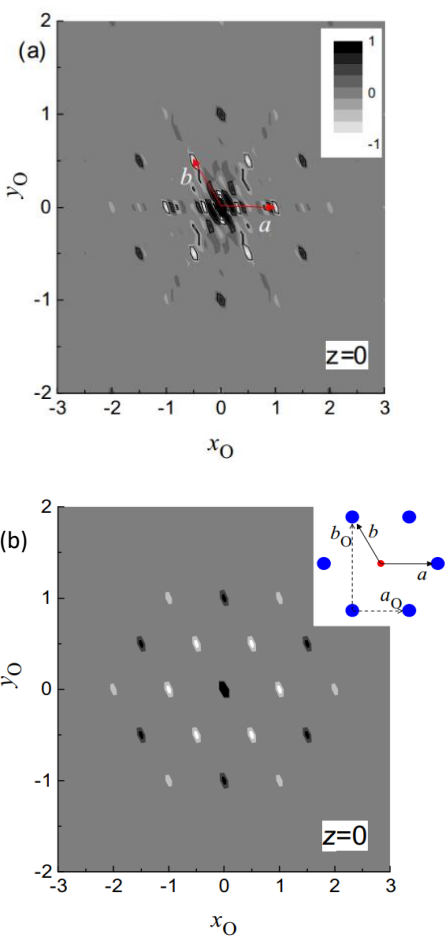

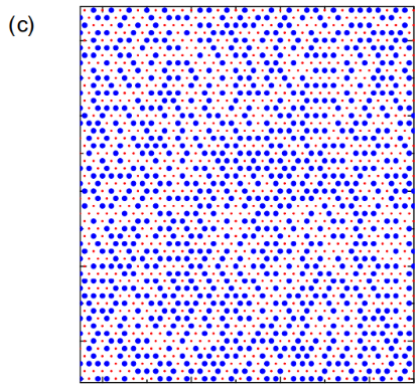

Figure 7. (a) Pair distribution function (PDF) within the $z=0$ plane derived from the measured diffuse X-ray scattering intensity distribution. The diffuse X-ray scattering was measured at room temperature using an orthorhombic sample because the hexagonal crystal was smaller than the orthorhombic one, whereas their intensity distributions were basically the same. (b) Calculated PDF for the cluster shown in the inset. (c) Real space Cu-Sb arrangement that well reproduces the PDF presented in (b). Difference in symbol size shows difference in element. Figure is taken from ref. [34].

- **Theoretical models of the crystal structure**

To understand the mechanism leading to the spin-orbital liquid state, it is important to further clarify the underlying crystal structure made by Cu ion. For this purpose, Monte Carlo simulations for a frustrated model of charged dumbbells on the triangular lattice has been performed [35], and showed a nontrivial lattice structure consistent with the X-ray diffraction results obtained for $Ba_3CuSb_2O_9$ [24]. It is found that a fractal branching structure with a fractal dimension $d_f = 1.90$ at $T = T_{frz}$ (a freezing temperature) is formed by the super-exchange-linked $Cu^{2+}$ clusters, indicating the presence of orphan spins [35]. The nearest-neighbor singlet model supports the scenario that the correlated dumbbell disorder promotes a spin-orbital liquid state with delocalized orphan spins [35]. Furthermore, density field theory has been used to further analyze the arrangement of the dumbbells, obtaining the conclusion that BCSO dumbbells are disordered with stripe-like arrangements at local level [36]. The structure with the lowest energy has a stripe arrangement of the dumbbells in each layer and stripes in neighboring layers are parallel to one another [36], forming a large gap of ~ 0.5 eV at the Fermi level. This makes it difficult for correlation effects to drive a dynamic JT distortion in the idealized crystal with a perfect stripe order [36]. Therefore, it is likely that the disorder plays an important role to produce the honeycomblike lattice, namely, randomly depleted honeycomb based lattice with the stripe order as shown in Figure 7c and to stabilize the spin-orbital liquid state, as proposed by A. Smerald and F. Mila [35].

**Two Structural Phases of BCSO**

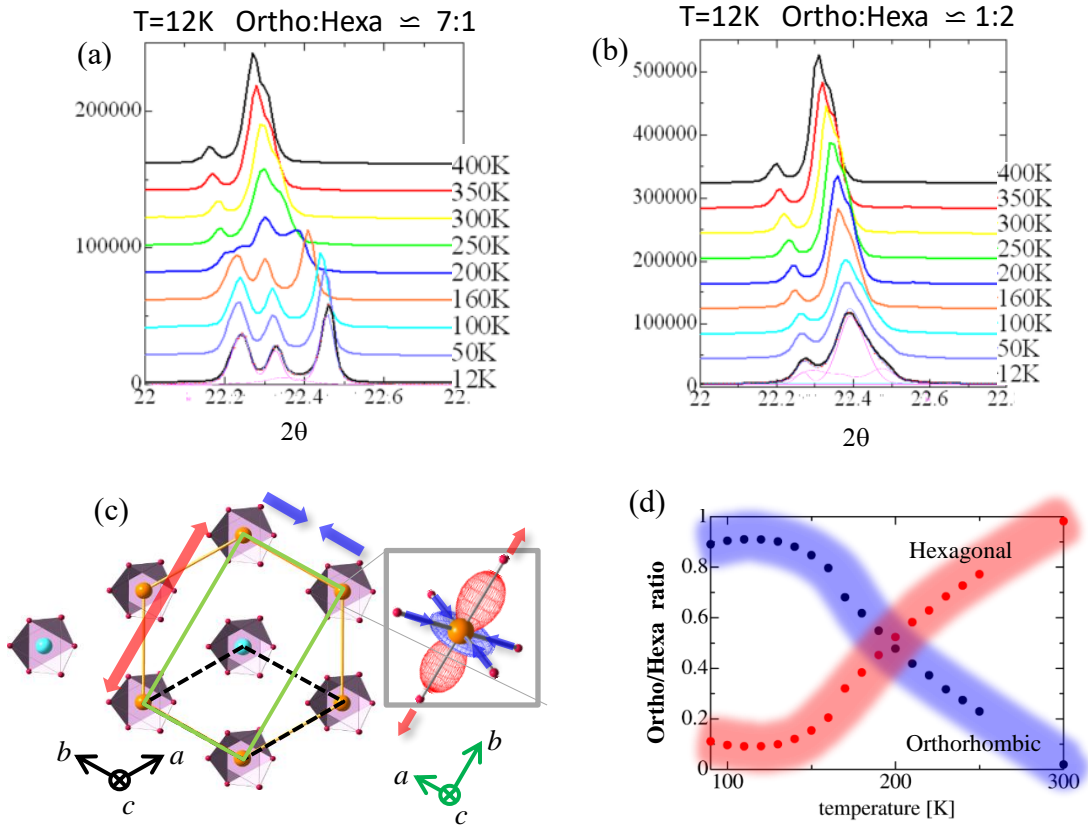

Figure 8. Powder X-ray diffraction pattern at several temperatures for (a) orthorhombic and (b) hexagonal samples. (c) Structural transition from hexagonal to orthorhombic sample is associated with the orbital order induced by cooperative distortion. (d) Phase volume of the hexagonal and orthorhombic structure as a function of temperature. Crossover occurs over a wide range of temperature.

As we discuss below, the existence of two distinct types of BCSO, namely, the hexagonal and orthorhombic phases, has allowed us to make detailed comparative study on the JT distortion and its effects on the magnetism as well as the orbital degrees of freedom. Thus, before discussing the magnetic properties and orbital ordering, we will first describe the structural difference in the two phases. The Jahn-Teller effect in BCSO causes $CuO_6$ octahedral distortion. The EXAFS measurements have been performed and indicated that all the Cu sites in both the hexagonal and orthorhombic samples have the same type of local JT distortions at least in the short time scale of $10^{-12}$ sec [24]. Namely, the EXAFS measurements have revealed the existence of two long Cu-O bonds (2.26 Å) and four short Cu-O bonds (2.03 Å) regardless of hexagonal or orthorhombic samples [24]. Orthorhombic lattice distortion occurs when the long-range ferro-orbital arrangement is established, while long-range antiferro-orbital arrangement should give rise to a superstructure. In order to examine the ferro-orbital arrangement, we investigated the temperature dependence of the powder X-ray diffraction patterns for two different types of sample. Figure 8 (a) shows that the system exhibits a structural phase transition that lowers the structural symmetry on cooling below 200 K. Based on the powder XRD structure analysis, we found that the space group is *P*6$_3$/*mmc* at 300 K, but about 94% of the sample undergoes the structural phase transition into the low temperature, orthorhombic, *Cmcm* phase, which has also been confirmed by the Raman scattering study [37]. A splitting of doubly degenerate phonons and a change of polarization dependence of the *A$_g$* phonons were observed from the Raman scattering spectra, indicating the structural changes due to the formation of the orthorhombic BCSO [37]. When the hexagonal structure becomes distorted into the orthorhombic structure at low temperatures, the three-fold axis of the dumbbell is lost, lifting the orbital degeneracy. Therefore, the observed orthorhombic distortion should be considered as the result of the cooperative JT effect (Figure 8c). As shown in Figure 8 (d), the volume ratio between the orthorhombic and hexagonal phases changes gradually with temperature from 300 K down to 150 K.

The consideration on the electronic state of CuSbO$_9$ cluster forming the Cu-Sb dumbbell provides the key to understand the reason why this phase transition occurs gradually over a wide temperature range. The Cu ions are JT-active while Sb$^{5+}$ is a spherical ion due to the absence of orbital degrees of freedom and must have ionic rigid bonds with the nearest neighbor oxygen anions based on the large difference in the ionic charge. Such structural rigidity of the SbO$_6$ octahedra competes against the JT effect of the CuO$_6$ octahedra to lower the energy gain by the JT distortion because Cu and Sb share a face in this CuSbO$_9$ cluster. Such competition or the "Sb capping effect" should suppress the transition from hexagonal to orthorhombic phase by orbital order, leading to the crossover type structural change over a wide temperature range. K. V. Shanavas *et al.* have studied the JT effect with the Coulomb correlated density-functional calculation for a hexagonal lattice, suggesting that the correlation effects are important for the JT distortions of the CuO$_6$ octahedra [38]. Meanwhile, they argue for a random static JT effect for the hexagonal structure. The broken spatial symmetry of the CuO$_6$ octahedra leads to one minimum energy level 43 meV lower than the other two, which is impossible for tunneling, then suggesting a static JT effect. The Cu-Sb flipped dumbbells will induce significant disorder in the system, resulting in random JT distortions [38]. However, this method may overestimate the localization energy while the spin-orbital resonance as well as the above "Sb capping effect" in the Cu-Sb dumbbell may enhance the quantum tunneling energy scale for orbital fluctuations.

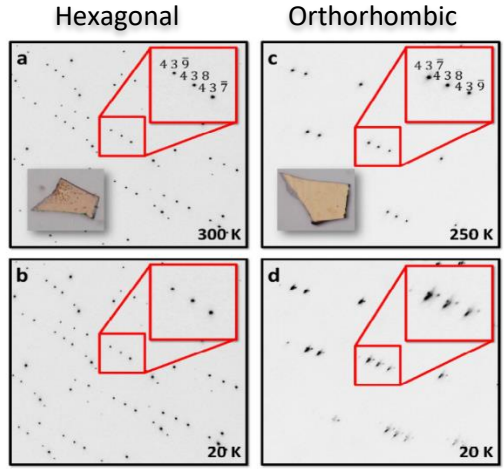

Figure 9. Single crystal X-ray diffraction of hexagonal samples, which has no transition at low temperature (a, b), and orthorhombic samples, which undergo hexagonal to orthorhombic structural transition at low temperature (c, d).

It is worth mentioning that although the orthorhombic phase is expected to be stabilized by the ferro-orbital order below the transition temperature, X-ray pattern indicates that several percent of component maintains three-fold axis even at the lowest temperatures. This fraction of the hexagonal phase in the orthorhombic sample at the lowest temperature is different from sample to sample. As shown in Figure 8 (b), there is one sample with two-thirds part remains undistorted down to the lowest temperature [24]. Recently, through the intensive study on the single crystal growth, we have succeeded in growing single crystals in which the phase transition does not occur down to the lowest temperatures (Figure 9). Unlike the case for the orthorhombic sample where diffraction peak splitting is observed due to the domain formation, the hexagonal sample retains the same symmetry in the reciprocal space as its high temperature phase. This hexagonal sample also possesses no cooperative Jahn-Teller distortion, as confirmed by several experiments including the X-band ESR measurement [39] and Raman spectra [37]. As there is no sharp superlattice reflection observed for hexagonal single crystals, no long-range antiferro-orbital ordering has been found in BCSO. Thus, our experiment clarifies that there are two phases in this system. A recent powder neutron diffraction study with a constant wavelength of 1.1969 Å and three orders shorter time scale than ESR [40] for $Ba_3Cu_{1+x}Sb_{2-x}O_9$ (x = 0, 0.1) indicated ferro-orbital order for x = 0.1 while no long-range orbital ordering for x = 0 [41], even though the $Cu^{2+}$ ions are JT distorted for both compositions. This is consistent with the results obtained by the extensive X-ray diffraction study showing that the stoichiometric single crystal remains hexagonal down to the lowest temperatures while the off-stoichiometry leads to the orbital ordering and stabilizes the orthorhombic phase [39].

To examine the possibility of the short-range arrangement of the orbital degree of freedom in hexagonal samples, diffuse X-ray scattering was measured [34, 42]. Each one of the JT distorted $CuO_6$ octahedra gives rise to a strain field in the crystal. Such a strain field provides a diffuse X-ray scattering around reciprocal lattice points, known as Huang scattering. While isolated point sources of distortion (in this case, JT distorted $CuO_6$ octahedra) lead to ordinary Huang scattering, the spatial correlation between the sources may produce the modulation of the diffuse intensity distribution. Figures 10 (a) and (b) show the diffuse X-ray scattering intensity distribution around 220 Bragg reflection measured at 290 K and 4 K. At both temperatures, the intensity distribution extends circumferential direction around the origin of the reciprocal space. Panels (c) and (d) show the calculated Huang scattering intensity distribution induced by the octahedra distorted in the JT mode and breathing mode, respectively. The experimentally observed diffuse intensity was found to be related to the JT distortion, and therefore, to the orbital degree of freedom.

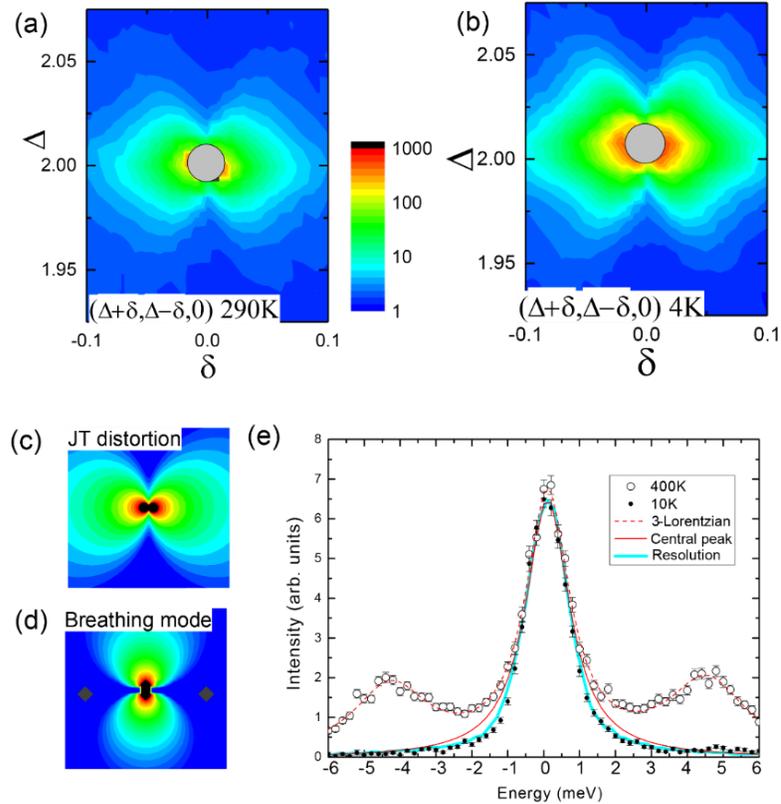

Figure 10: (a) Intensity distribution around the 220 Bragg reflection measured at 290 K. (b) That at 4 K. (c) Calculated Huang diffuse intensity induced by a JT mode distortion. (d) That by a breathing mode distortion. All the maps present the intensity in a log scale. Figures are taken from [42]. (e) Inelastic X-ray scattering spectrum of the diffuse intensity measured at (4.2 3.8 0) at 400 K (open symbols) and 10 K (closed symbols) normalized at zero energy transfer. Three-Lorentzian fitting was used to extract the central peak for 400 K data. The resolution profile is also presented. Figure is taken from ref. [34]. The experimental data of (a)-(e) are all measured for the hexagonal samples.

**Magnetic Properties: Formation of a Gapless Spin Liquid State at Low $T$s**

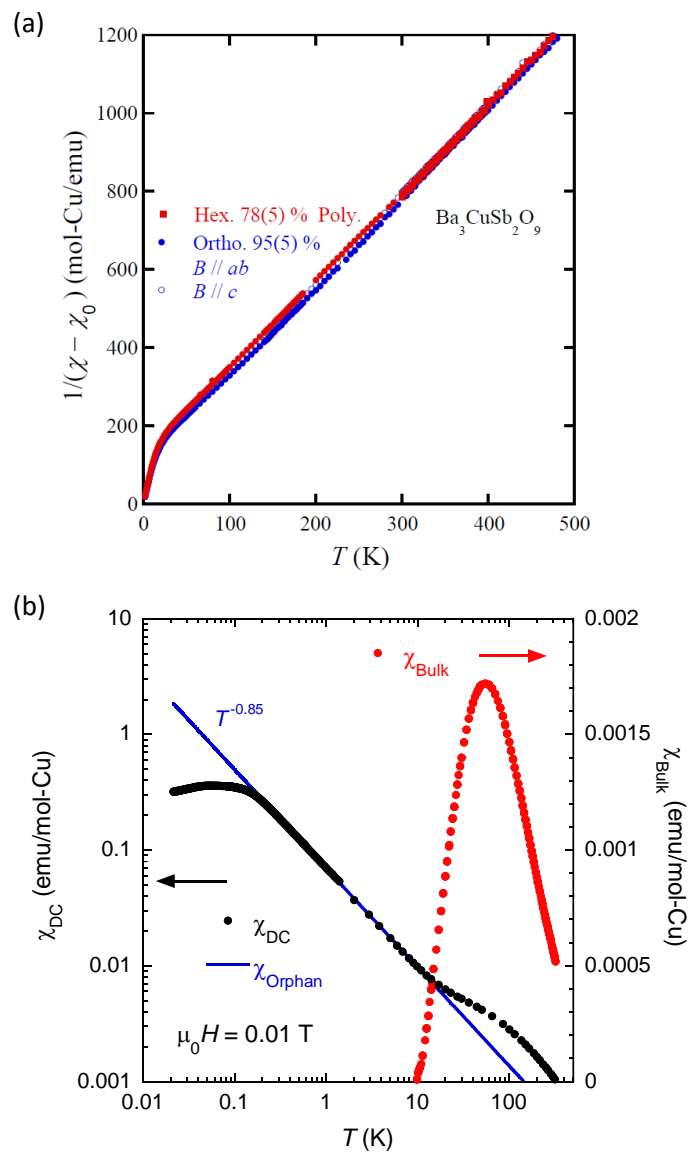

Figure 11. Temperature dependence of the magnetic susceptibility for a single- and polycrystalline samples of $Ba_3CuSb_2O_9$ in a magnetic field of 100 Oe applied in the *ab* plane. (a) Temperature dependence of the inverse magnetic susceptibility. Here, "Hex. 78(5)% Poly." denotes a polycrystalline sample with 78(5)% volume fraction of the hexagonal phase at low temperatures, while "Ortho. 95(5)%" indicates a single crystal with 95(5)% fraction of orthorhombic phase at low temperatures. The phase fractions are obtained from the refinement of the X-ray diffraction spectra. (b) Magnetic susceptibility for the hexagonal sample at low temperature which can be decomposed into two components, namely the bulk spin and orphan spin components.

To study the magnetic properties of the system, we performed the magnetic susceptibility measurements. At higher temperatures, we observed the Curie Weiss law consistent with antiferromagnetically interacting $S = 1/2$ spin systems, as expected for a lattice made of $Cu^{2+}$ ion. The Weiss temperature corresponds to AF coupling with the strength of 45 K. By taking into account the coordination number, the super-exchange interaction of the nearest neighboring Cu is estimated to be about 50 K. At lower temperature, a deviation from Curie Weiss law was observed. The downturn below 50 K in the inverse susceptibility versus temperature curve (Figure 11a) indicates the presence of orphan spins, which may originate from the structurally isolated Cu ions (Cu' spins). Therefore, the total magnetic susceptibility can be thought as the sum of two components, the bulk honeycomb Cu spins on Cu sites and the orphan Cu spins on Cu' sites. A log-log plot of the magnetic susceptibility against temperature reveals a systematic increase on cooling with a hump around 50 K. A power law fit $\chi = T^{-\alpha}$ yields the coefficient α of ~ 0.85. Subtracting the total magnetic susceptibility with this power-law component, we obtained the susceptibility component that peaks at 50 K (Figure 11b). This component most likely comes from the bulk honeycomb lattice Cu sites as suggested by the neutron scattering experiments performed at National Institute of Standards and Technology, which found a singlet dimer type of short-range correlation on the honeycomb based lattice between the local neighboring Cu ions (Figure 12).

J. Nasu *et al.* have constructed the quantum dimer model made of spin-orbital singlet dimer and local JT singlet [26]. Spin-orbital singlet dimer is that two spins in a nearest-neighbor bond form a singlet state with the orbital polarization along the bond. Local JT singlet is a state where the dynamical JT effect leads to quenched orbital polarization, as indicated by the ESR experiments with almost isotropic *g* factor down to 30 K, suggesting no specific orbital alignments but a possibility of orbital quenching [24, 39, 40]. Both the spin-orbital singlet dimers and the local JT singlets can hop quantum mechanically due to a competition between the super-exchange interaction and the dynamical JT effect, which has been proposed to be a candidate state [26, 43] for the non-magnetic-ordered phase observed in $Ba_3CuSb_2O_9$. The power law behavior, on the other hand, is one of the signatures for the system with random singlet ground states. In random singlet states, the spins form singlet pairs with each other randomly and the pairing distance is determined by the correlation length, which tends to increase with decreasing temperature [24]. A Raman spectroscopy study combined with lattice dynamics calculations supports a coexistence of spin-orbital liquid and random singlet state in $Ba_3CuSb_2O_9$ [44].

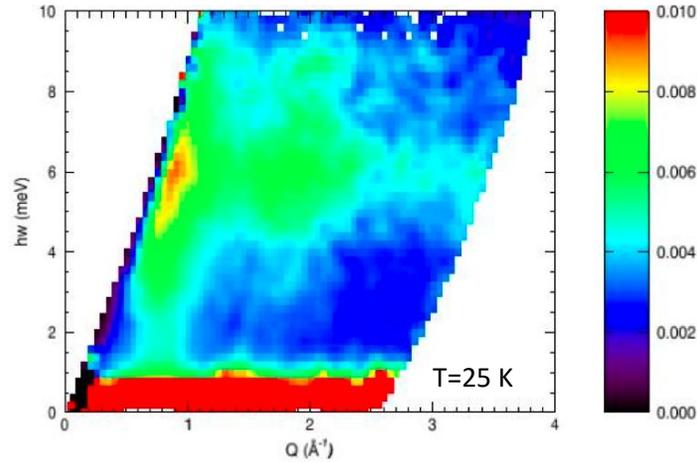

Figure 12. Spin liquid state forming an energy gap revealed by neutron scattering in the hexagonal sample at 25 K. Vertical axis represents energy, while horizontal axis represents reciprocal wave number. The gap formation stabilizes the spin liquid state and makes it robust against impurity.

In low temperature region, AC and DC susceptibility measurements on hexagonal single crystal show a peak associated with spin freezing at 120 mK. This spin freezing is found to be glassy in nature, as indicated by the hysteresis between zero-field cooled and field cooled DC susceptibility, as well as the slight frequency dependence of the freezing temperature in the AC susceptibility. Moreover, the susceptibility below freezing temperature does not drop rapidly to zero, suggesting that only a minor part of spins is frozen, consistent with the dynamical spin liquid state confirmed by the muon spin resonance technique (Figure 14) [24, 45].

Figure 13 (a) shows the temperature dependence of the specific heat. The lattice contribution was estimated by the specific heat obtained for the nonmagnetic analog $Ba_3ZnSb_2O_9$ carrying no magnetic elements. A Schottky-type anomaly is observed in the magnetic specific heat part below 10 K. The hump associated with the Schottky anomaly is gradually suppressed as applied magnetic field increases, but remains visible under magnetic fields up to 9 T. Thus, the magnetic specific heat can be regarded as the sum of the bulk specific heat from Cu spins on Cu sites, and the Schottky specific heat from orphan spins.

The Schottky specific heat is modeled by using a two-energy-level (single gap) model and subtracted from magnetic specific heat to obtain the bulk specific heat. It is interesting to note from Figure 13 (b) that the bulk specific heat under magnetic field greater than 5 T shows a linear $T$ dependence and reduces to a non-zero γ-value as the temperature goes to zero. This behavior is consistent with previous study on polycrystalline sample by H. D. Zhou *et al*. However, the bulk specific heat for the magnetic field below 5 T cannot completely be extracted by single gap Schottky anomaly as described above. At least two gap values are required to remove Schottky contributions completely, as shown in Figure 13 (c). This multiple gap value is consistent with the recent NMR study [45] which describes the crossover to the state with distributed gap below 10 K.

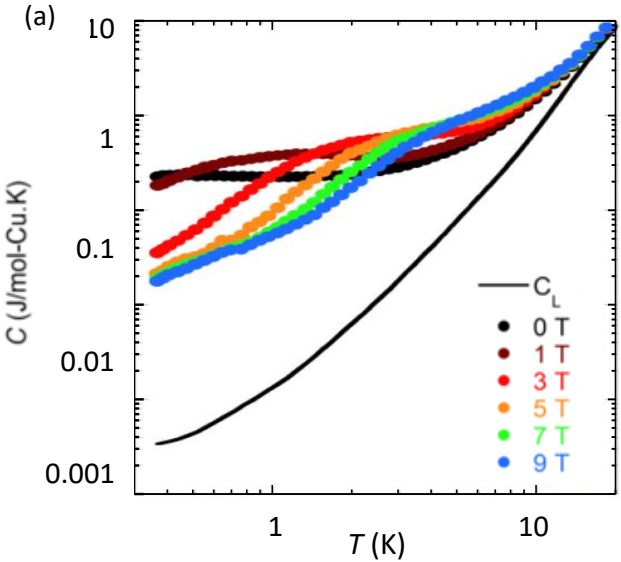

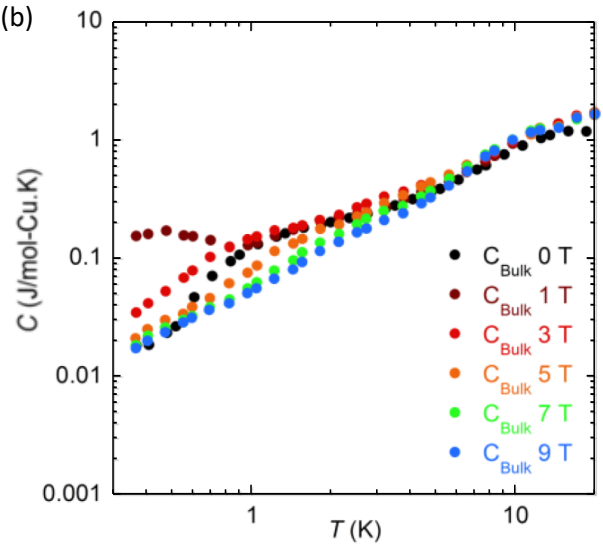

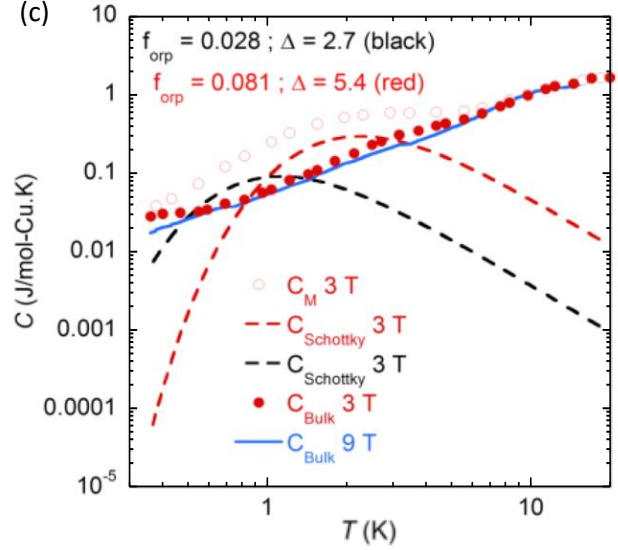

Figure 13. Temperature dependence of the specific heat for a hexagonal sample. (a) Total specific heat as a function of temperature under various magnetic fields. $C_L$ represents the lattice contribution estimated using the analog $Ba_3ZnSb_2O_9$. (b) Temperature dependence of the bulk specific heat $C_{Bulk}$ at various magnetic fields after the subtraction of the single gap Schottky specific heat. (c) Schottky analysis for the specific heat at 3 T. Magnetic part $C_M$ (open circle) is estimated by subtracting $C_L$. Bulk specific heat $C_{Bulk}$ at 3 T (closed circle) is estimated by subtracting the two respective Schottky anomalies $C_{Schottky}$ (black and red broken lines) with different gap values $\Delta$ = 2.7 and 5.4 K and weight $f_{orp}$ = 0.028 and 0.081. Interestingly, the bulk specific heat at 3 T obtained from the above procedure overlaps on top of the bulk specific heat at 9 T (blue solid line) shown in the panel (b).

In both low temperature magnetic susceptibility and specific heat, the orphan spin contribution dominates at low temperatures, allowing to estimate the amount of orphan spins by low temperature analyses. In the case of magnetic susceptibility, the ratio between the effective moment size obtained by the Curie-Weiss fit in the low temperature and near-room temperature regions, tells the number of orphan spins, given that at near-room temperature the entire magnetic susceptibility comes from the bulk spin (Cu spins on Cu site). In our single crystalline samples, this orphan spin amount is estimated to be 12% for the hexagonal sample. In the case of specific heat, the magnetic specific heat at a given applied magnetic field can be fitted with the Schottky anomaly. This analysis yields the maximum value of 12% orphan spin in the hexagonal single crystal.

As discussed above, the bulk specific heat exhibits a linear temperature dependence $C \approx \gamma T$, which is particularly noticeable under high magnetic fields. This linear temperature dependence is characteristic for the specific heat of various quantum spin liquid candidates, such as $\kappa$-(BEDT-TTF)$_2$Cu$_2$(CN)$_3$ [46], EtMe$_3$Sb[Pd(dmit)$_2$]$_2$ [47] and Na$_4$Ir$_3$O$_8$ [48]. For BCSO case, we obtained Sommerfeld constant $\gamma \sim 50$ mJ/mol$^{-2}$K$^{-2}$, which is in the range between 1-250 mJ/mol$^{-2}$K$^{-2}$ found in the spin liquid candidates. This Sommerfeld constant indicates the density of low-energy states. This is consistent with the distribution of the spin gap as revealed by neutron scattering measurements [24] and further suggests that the randomness plays an important role. In this context, K. Uematsu *et al.* have recently studied the bond-random $S = 1/2$ Heisenberg model on a honeycomb lattice, indicating the randomness induces the gapless quantum spin liquid (QSL) state, which might be applicable to Ba$_3$CuSb$_2$O$_9$ [49].

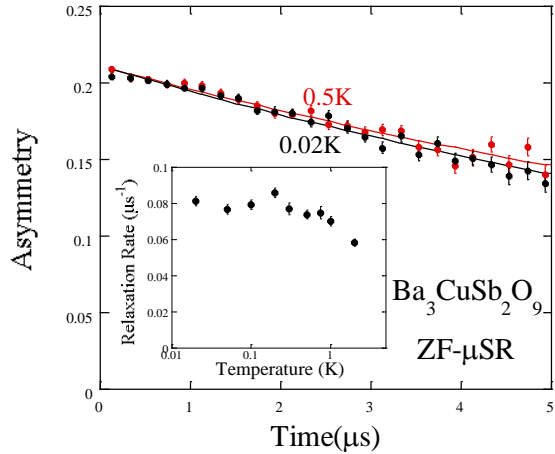

Figure 14. Zero-field μSR spectra for a hexagonal sample exhibiting a single-exponential decay exp(-λt). Inset shows the temperature dependence of relaxation rate λ.

Moreover, to further investigate the spin dynamics, we performed the muon spin resonance (μSR) experiments to estimate the internal magnetic field. If a magnetic order occurs, a time dependent oscillatory signal should be seen in the muon spin polarization asymmetry. However, as shown in Figure 14, the relaxation at 20 mK and 0.5 K occurs very slowly without any wiggle. This indicates that the Cu spins fluctuate in the time scale much faster than that of the muon decay time scale (μsec). Additionally, as shown in inset of Figure 14, the relaxation rate at low temperature from 1 K down to 20 mK barely changes, indicating the absence of any spin freezing at 110 mK, where the AC and DC susceptibilities show an anomaly. This provides strong evidence for quantum spin liquid state as a rare example not only among copper oxides but also in the insulating magnetic materials.

To further reveal the BCSO's ground state, both nuclear magnetic resonance and μSR measurements have been performed by J. A. Quilliam *et al*. [45] for polycrystalline samples of BCSO. While μSR results confirmed the absence of magnetic order down to 20 mK consistent with the above observation made in ref. [24], NMR measurement revealed a non-monotonic shift in the $^{123}$Sb line spectra with peak around 55 K, consistent with our inelastic neutron scattering and magnetization results indicating the gap size of 50 K (Figures 11 and 12). The comparison between the temperature dependence of the shift and the susceptibility approximated by a high-temperature series expansion fit for the $S = 1/2$ two dimensional triangular antiferromagnet indicates that the BCSO magnetic excitation is not similar to the 2D triangular AFM. Indeed, the temperature dependence of the intrinsic susceptibility derived from the NMR line shift, however, is approaching zero as temperature goes to 0 K, which is quite different from the finite susceptibility found in several quantum spin liquid candidates. Although the sample was assigned as hexagonal in the paper, the field dependence of the magnetization, and the stoichiometry of Cu and Sb is more consistent with the orthorhombic sample based on our detailed study [39]. Thus, this gap distribution may be attributed as the ground state of orthorhombic BCSO, a singlet state with static orbital order. Besides, the power-law *T*-dependent transverse thermal conductivity [50] indicates the spin excitation is gapped, consistent with the neutron and NMR measurements [24, 45]. Very interestingly, phonon Hall effect has been observed due to strong phonon scattering of orphan $Cu^{2+}$ spins [50].

**Evidence for the Orbital Liquid State in the Hexagonal BCSO**

In previous section, we have determined the space group for $Ba_3CuSb_2O_9$ to be *P*6$_3$/*mmc* hosting the short-range order of the Cu honeycomblike lattice. Although the space group is different from the previous reports [29, 30], hexagonal BCSO possesses $Cu^{2+}$ orbital degrees of freedom irrespective of space group, in contrast to orthorhombic BCSO. Materials with degenerate orbital degrees of freedom under a cubic crystal electric field are known to undergo a structural transition inducing Jahn-Teller distortion at low temperatures. In addition, it is common that a magnetic order arises from exchange interactions determined based on the ordered configuration of orbitals. In particular for the honeycomb based lattice (bipartite lattice) the nearest neighbor antiferromagnetic exchange coupling is known to lead to an antiferromagnetic order. Therefore, there should be a non-trivial reason why $Ba_3CuSb_2O_9$ does not show any magnetic order or spin freezing.

Temperature variations of the ferro- and antiferro-correlation between the orbitals are estimated from the diffuse intensity distribution as presented in Figures 15 (a) and (b), respectively. As can be seen, both correlations develop with decreasing temperature down to 60 K. The saturation of the orbital correlation around 60 K implies the interaction between the orbital and spin degrees of freedom, because the exchange coupling scale and the spin gap size found in the bulk Cu sites in BCSO is close to this temperature (Figures 11 and 12).

Interestingly, as we discuss below, the ESR measurements find a quantum orbital disordered state develops below ~ 40 K, indicating the formation of a spin-orbital entangled liquid state [40].

To study the time scale of the orbital motion, we performed inelastic X-ray scattering measurements for the Huang scattering. The result is presented in Figure 10 (e). While the quasielastic signal measured at 10 K is as narrow as the instrumental resolution, the signal measured at 400 K is slightly broader than the instrumental resolution. The sharp profile observed at low temperature means the time constant of the lattice distortion providing the Huang scattering is much slower than the instrumental resolution (~ 3 ps), at least at low temperatures. The 0.15(4) meV broadening in the FWHM of the quasielastic component observed at 400 K, compared to the resolution profile may be caused by the short lifetime of orbital motion. Another possible origin is multi-phonon scattering, sometimes giving a broad peak around zero-energy transfer [34]. The orbital motion in BCSO studied by the Huang scattering is found to be rather slow, a few picoseconds or slower [34], consistent with the recent report of ESR measurements as described in the next paragraph [40].

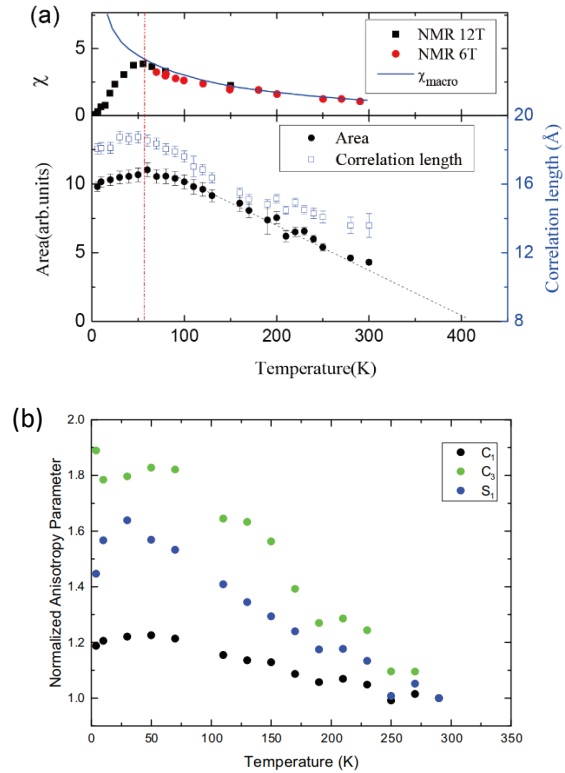

Figure 15: Temperature variation of the orbital correlation in the hexagonal BCSO. (a) Ferro-orbital order parameter (solid circle) and correlation length (open squares) together with the macroscopic magnetic susceptibility (solid line) and Knight shifts measured at magnetic fields of 12 T and 6 T reported in Quilliam *et al* [45] (black squares and red circles). Figure is taken from [42]. (b) Antiferro-orbital order parameter $C_3$ as a function of temperature. Figure is taken from ref. [34].

Significantly, the extensive studies made by Y. Han *et al.* have revealed that the orbital fluctuations develop a quantum orbital disordered state with a nearly constant time scale of ~ 100 ps at low temperatures below 40 K in the stoichiometric single crystalline sample of the hexagonal phase, in sharp contrast with the static JT effect and orbital ordering observed in the orthorhombic phase [40]. S.-H. Do *et al.*'s investigations of hexagonal and an orthorhombic majority phases of BCSO through high-frequency ESR [51] fully agree with these results. ESR spectroscopy serves as a key technique to delineate the orbital dynamics at the bulk Cu sites. The dynamic orbital fluctuations would lead to a crossover from an isotropic to a static anisotropic line shape with increasing frequency according to two mechanisms. (I) Snapshot effect: when $1/\nu_{EM} \gg \tau$, where $\nu_{EM}$ is the electromagnetic wave frequency and $\tau$ is the time scale of orbital tunneling, an isotropic ESR line shape could be observed. Conversely, when $1/\nu_{EM} \ll \tau$, magnetic molecules would be in a very slow motion. A static JT distortion would give rise to a static anisotropic ESR line shape. (II) Decreasing temperature, increasing internal strain, or increasing Zeeman energy would quench orbital quantum fluctuations. A static JT distorted ESR spectra would be expected when the anisotropic Zeeman energy is much larger than the orbital tunneling energy. Figures 16 a-d show the frequency dependence of the line shape of the ESR absorption curves for two different field directions as shown in the insets of figures, illustrating that the spectra become deformed from a Lorentzian curve when the frequency exceeds a critical value: $\nu_C$ ~ 80 GHz at 1.5 K (Figures 16a, b), and $\nu_C$ ~ 180 GHz at 50 K (Figures 16c, d). Meanwhile, the line shape at 730 GHz for the hexagonal BCSO looks similar to that in the orthorhombic BCSO at 9.3 GHz, exhibiting a high frequency "snapshot" of the dynamic JT effect. Notably, the single Lorentzian line width decreases with increasing frequency at $\nu_{EM} < \nu_C$, as a result of low frequency "snapshot". $\nu_C$ shifts to higher frequencies with increasing temperature due to that thermal fluctuations speed up the orbital tunneling. Both the observations are opposite to the tendency expected for the exchange splitting effect. In addition, the periodicity in the angle dependence of the g-factor dramatically changes at ~ 100 GHz as shown in Figures 16e-h. Looking down from the [001] direction, three equivalent O-Cu-O bonds that could elongate due to the Jahn-Teller distortion, and the directions of the bonds projected in the c-plane form 120 degrees to each other. Therefore, we have three sets of two-fold periodic angular dependent signals in Figure 16 (g) and (h). Below 100 GHz, almost isotropic *g*-factors with the tiny six-fold periodicity (note the vertical scale) are observed, while three sets of two-fold periodic *g*-factors are observed above 100 GHz (Figure 16g), which is similar to those observed at 9.3 GHz for the orthorhombic sample (Figure 16h). Thus, these observations provide strong evidence for the dynamic orbital fluctuations.

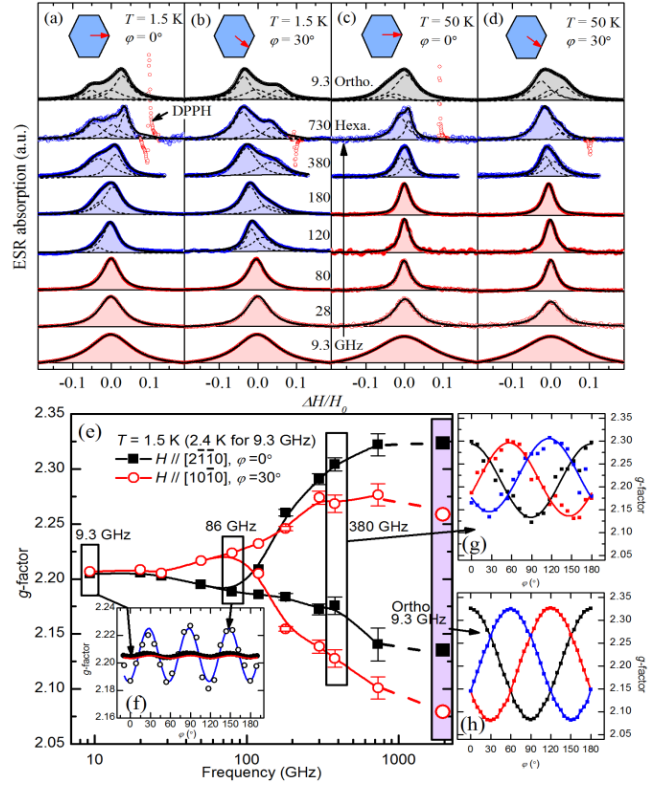

Figure 16. (a)-(d) ESR absorption spectra of various electromagnetic wave frequencies $\nu_{EM}$ with multipeak Lorentzian fits and the total fits. (a) $\varphi = 0°$, $T = 1.5$ K; (b) $\varphi = 30°$, $T = 1.5$ K; (c) $\varphi = 0°$, $T = 50$ K; (d) $\varphi = 30°$, $T = 50$ K. $\nu_{EM}$ varies from 9.3 to 730 GHz for hexagonal BCSO, while $\nu_{EM}$ = 9.3 GHz for orthorhombic BCSO. $\Delta H / H_0 = (H - H_0)/H_0$ , $H_0 = h\nu_{EM}/(g_1\mu_B)$, $g_1 = 2.2$. (e) Frequency dependence of $g$-factors for $H//[2\bar{1}\bar{1}0]$ ($\varphi = 0°$) and $H//[10\bar{1}0]$ ($\varphi = 30°$) for the hexagonal sample and those for the orthorhombic sample (rightmost data points). The inset figures indicate the angular dependences of the $g$-factors at (f) 9.3 GHz and 86 GHz, (g) 380 GHz for the hexagonal sample and (h) 9.3 GHz for the orthorhombic sample.

Further investigation on the local structure and the temperature dependence of the JT distortion has been performed with Raman spectroscopy and the ultrasound measurements. Differences between the hexagonal and orthorhombic samples are observed in the temperature dependence of the elastic constants obtained by the ultrasound measurements [39]. For the orthorhombic sample, an elastic softening was observed in $C_{44}$ near the phase transition, corresponding to the freezing of $d_{x^2-y^2}$, $d_{y^2-z^2}$, and $d_{z^2-x^2}$ orbitals. A Curie-Weiss-like-fit based on the theoretical elastic constant for the cooperative JT effects $C_\Gamma(T) = C_{\Gamma 0}(T) \frac{T - T_C}{T - \Theta}$ yields the Curie-Weiss constant of $\Theta \sim 250$ K, indicating the ferro-orbital correlation [39]. In the hexagonal sample, no characteristic softening was observed down to the lowest temperature [39]. From Raman spectra, both samples show high disorder and local broken symmetry in the edge-sharing octahedra, which leads to a split in the shared-face oxygen mode near 550 cm$^{-1}$. The spectra for the hexagonal sample are unchanged on cooling, whereas a new band at 587 cm$^{-1}$ is found for the orthorhombic sample, indicating the symmetry lowering associated with the JT distortion [39]. Further Raman spectroscopy measurements by N. Drichko *et al*. [37] with the time scale of ~ ps also demonstrate that the hexagonal BCSO preserves the average hexagonal symmetry down to 20 K, while the orthorhombic BCSO shows a transition from a hexagonal *P*6$_3$/*mmc* to an orthorhombic *Cmcm* structure below 200 K [37]. The analysis of the stretching oxygen vibrations suggests that local JT distortion of the Cu$^{2+}$ environment in the hexagonal sample increases on cooling but is always smaller than the distortion in the orthorhombic sample. The observation is fully consistent with the dynamical JT distortion with the time scale of ~ 10 ps [37].

Among two possible scenarios for the absence of the cooperative JT effect in the hexagonal single crystal even down to the lowest temperatures: (1) orbitals at each Cu site become totally frozen and orient randomly, forming an orbital glass state, or (2) orbitals dynamically change the quantized axes among the three possible directions, our observations clearly support the latter case. Namely, unlike the orthorhombic sample, freezing of orbital fluctuations is not found in the hexagonal samples down to 4 K, as confirmed by X-ray diffraction, diffuse X-ray scattering, inelastic X-ray scattering, Raman, and ultrasound measurements. The most significant observations have been made using ESR technique and have provided evidence for the dynamic JT state in the hexagonal sample (Figure 16). Moreover, when we consider the specific heat peak at 20 K, a larger magnetic contribution is observed for the hexagonal sample than for the orthorhombic sample [24]. This is consistent with the fact that the hexagonal sample retains the strong orbital fluctuations resonating with the spin fluctuations scale set by the pseudo spin gap energy of 40 K, while the orbital degree of freedom is frozen by the JT distortion in orthorhombic sample at much higher temperature of ~ 200 K. In addition, the magnetization and neutron scattering measurement results indicate that neighboring Cu ions form a ferro-orbital order that stabilizes a singlet dimer on the bond connecting the neighboring Cu sites. Thus, as orbitals do not freeze, the most likely case is that spin and orbital form a dynamic state, a quantum-mechanically resonant state. Given the comparable energy scale between spin interaction (*zJ*

~ 200 K) and orbital fluctuation (200 K), spin may correlate with orbital fluctuations, forming a spin-orbital liquid state (Figure 17).

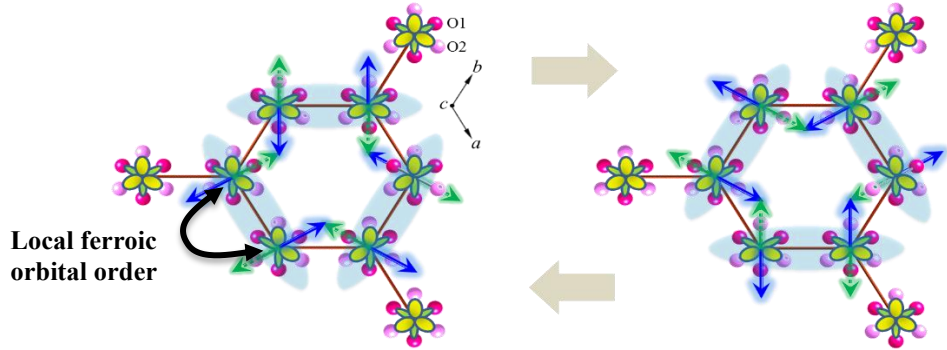

Figure 17. Possible quantum resonant state formed by spin-orbital entanglement in $Ba_3CuSb_2O_9$, which resembles the π electron resonant state in benzene. Arrows indicates the spins forming singlet dimers (a pair of blue and green arrows). The neighboring orbitals forms a local ferroic order, which moves around on the honeycomblike lattice of BCSO.

The question would be how the two distinct states, namely orbital liquid and solid, may respectively appear in BCSO system in the hexagonal and orthorhombic phases at low temperatures despite almost the same composition and structure. Theoretical point of view would provide two plausible limiting cases here to stabilize both orbital liquid and orbital order. On one hand, orbital order is expected as the ground state for a spin-orbital system on a triangular lattice [9]. On the other hand, spin and orbital form an entangled state stabilizing a liquid state down to low temperatures on the honeycomb lattice [52]. This dynamic ground state accompanied by the resonance on the Cu hexagon resembles the π electrons in benzene (Figure 17). These two limiting cases might be actually realized by the presence of the frustration of the Cu-Sb dumbbell, creating the mixture of the honeycomb and triangular lattices in the crystal. The fraction of these two lattices might be the tuning parameter to control the ground state between the orbital liquid and solid. We should note that both honeycomb and triangular lattices are highly frustrated lattice structures for the entangled degree of freedom between spin and orbital. So taking the BCSO system as an example, frustrated lattice should be an important component to search for another spin-orbital liquid candidate.

**Conclusion**

In this paper, we have reviewed a series of experimental and theoretical studies that have revealed the formation of a spin-orbital entangled liquid state in the copper oxide $Ba_3CuSb_2O_9$. In particular, we went through the striking results obtained in the structural and magnetic measurements using macroscopic and microscopic probes. The synchrotron X-ray diffraction investigation shows BCSO has the lattice space group of $P6_3/mmc$ rather than previously reported $P6_3mc$, together with the frustrated arrangement of the Cu-Sb dumbbell [24]. The resultant short-range order of the Cu-Sb dumbbell orientation is confirmed by the fluorescence X-ray holography and X-ray diffuse scattering in accordance with the EXAFS result [24, 34]. The two types of the structural phases were found by the study using X-ray, EXAFS and Raman scattering. Namely, it was discovered, as the first case in a Jahn-Teller (JT) active transition metal compound, the "hexagonal phase" of BCSO does not exhibit any JT transition, keeping the hexagonal symmetry down to the lowest temperatures. In contrast, the "orthorhombic" BCSO undergoes a JT phase transition forming an ferro-orbital order at low temperatures [24, 37]. The neutron diffraction and magnetic susceptibility measurements are consistent with the muon spin resonance technique, confirming a gapless spin liquid state [24, 30, 45]. While the susceptibility and specific heat show that the orphan spin contribution dominates at low temperatures, the neutron inelastic scattering reveals the existence of a bulk spin gap due to the formation of the spin singlet dimers using the neighboring Cu $S = 1/2$ spins [24]. Besides, ESR experiments provided strong evidence for the dynamical orbital liquid state [24, 39, 40], and a Raman spectroscopy study supports a coexistence of spin-orbital liquid and random singlet state in $Ba_3CuSb_2O_9$ [44]. The absence of the cooperative Jahn-Teller distortion and no freezing of orbital fluctuations in hexagonal samples are further confirmed by various techniques, such as ESR measurement [39], Raman spectroscopy [37], and ultrasound measurements [39]. It has been found that a quantum orbital disordered state develops, accompanied by orbital fluctuations at a time scale of ~ 100 ps below 40 K in the hexagonal sample [40]. Inelastic X-ray scattering measurements for the Huang scattering as well as ESR measurements reveal the interaction between the orbital and spin degrees of freedom below the energy scale of superexchange coupling. All the above experiments provide strong evidence for the formation of a spin-orbital entangled liquid state [40], where a local singlet dimer formed on a neighboring ferroic orbital order moves around on the lattice, taking advantage of the dynamical Jahn-Teller effect.

To date few candidates for orbital liquid had been proposed prior to BCSO. It is very surprising to find the absence of macroscopic JT distortion in an insulating copper oxide material and a novel quantum liquid: "orbital liquid state". This can be viewed as similar to the RVB state that was proposed for the high-$T_c$ superconductors [53], and furthermore, would be now realized in a special lattice system. In addition, such a novel spin-orbital state has emerged on a Cu based honeycomb lattice as a result of the self-organized arrangement of the Cu-Sb dumbbells through the interaction of the associated electric dipole. The future control of the ground state by changing the sequence of the electric dipole would provide important guidance for designing quantum materials.

The existence of the large structural disorder makes it more surprising to find such a novel quantum liquid state in BCSO. Till 1970s, a lot of substances had been synthesized, and their structural details are all now available in the crystal structure database. However, there may be still a lot of unknown phases with a new structural type as was the case for copper oxide superconductors, and there may be still some compounds many of whose physical properties may have been overlooked just like in this case of BCSO. One of the driving forces in the study of the solid-state physics is the discovery of novel states of matter. Deepening the understanding of such states based on the detailed structural analysis method will provide a key clue for the future discovery of new materials. We hope that following this research, the search for unique quantum state will be even more successful in the future.

**Acknowledgment**

We would like to express special thanks to our collaborators: K. Kuga (Institute for Solid State Physics, The University of Tokyo, Present address: Graduate School of Science, Osaka University), K. Kimura (Institute for Solid State Physics, The University of Tokyo, Present address: Department of Advanced Materials Science, The University of Tokyo), R. Satake, N. Katayama, and E. Nishibori (Engineering Department, Nagoya University), R. Ishii ( Center for Quantum Science and Technology under Extreme Conditions, Osaka University, Present address: Institute for Solid State Physics, The University of Tokyo) and Y. Han ( Center for Advanced High Magnetic Field Science, Graduate School of Science, Osaka University, Present address: Wuhan National High Magnetic Field Center, Huazhong University of Science and Technology), F. Bridges (California State University), S. Tsutsui (Japan Synchrotron Radiation Research Institute, SPring-8), A. Q. R. Baron (Materials Dynamics Laboratory, RIKEN SPring-8 Center, RIKEN), Y. Ishiguro and T. Kimura (Division of Materials Physics, Graduate School of Engineering Science, Osaka University), T. Ito and W. Higemoto (Advanced Science Research Center, Japan Atomic Energy Agency), Y. Karaki (Faculty of Education, University of Ryukyus), A. A. Nugroho (Bandung Institute of Technology), A. Rodriguez-Rivera, M. A. Green (The United States National Institute of Standards and Technology), C. Broholm (Johns Hopkins University), T. Nakano and Y. Nozue (Graduate School of Science, Osaka University, TN's Present address: Ibaragi University). We would also thank for the collaboration with Y. Uesu, J. Kaneshiro and H. Yokota (Waseda University) for SHG measurement. This work is partially supported by PRESTO and CREST, Japan Science and Technology Agency (No. JPMJCR15Q5), and Grants-in-Aid for Scientific Research (No. JP25707030, JP24244059, JP26105008), and Program for Advancing Strategic International Networks to Accelerate the Circulation of Talented Researchers (No. R2604) from the Japanese Society for the Promotion of Science, and by Grants-in-Aids for Scientific Research on Innovative Areas (15H05882, 15H05883) of the Ministry of Education, Culture, Sports, Science, and Technology of Japan.